# Effect of Anode Dielectric Coating on Hall Thruster Operation

L. Dorf, Y. Raitses, N. J. Fisch

Princeton Plasma Physics Laboratory, Princeton, NJ, 08543

V. Semenov

Institute of Applied Physics, Nizhny Novgorod, Russia

An interesting phenomenon observed in the near-anode region of a Hall thruster is that the anode fall changes from positive to negative upon removal of the dielectric coating, which is produced on the anode surface during the normal course of Hall thruster operation. The anode fall might affect the thruster lifetime and acceleration efficiency. The effect of the anode coating on the anode fall is studied experimentally using both biased and emissive probes. Measurements of discharge current oscillations indicate that thruster operation is more stable with the coated anode.

In a gas discharge, there can be either an increase or a drop in the plasma potential toward the anode, generally referred to as the "anode fall". When the anode is at a higher potential than the near-anode plasma, we call the anode fall "positive", and when it is at a lower potential – "negative". In a Hall thruster discharge (HT),[1] the anode fall might affect the overall operation of the device. In the case of a positive fall, in which electrons gain kinetic energy in going from the plasma to the anode, the electron energy flux toward the anode is higher than in the case of a negative fall, when electrons are repelled. The increase in the power deposition near and at the anode might result in a decrease of the thruster acceleration efficiency[2], an increase of the anode heating (which may decrease the thruster lifetime), or additional ionization inside[3] or near[4] the anode.

In spite of a number of experimental[3-12] and theoretical[13-15] studies of a HT internal plasma structure, the understanding of the intra-anodal and near-anode processes in HTs is still very limited. It was suggested theoretically that steady-state operation of a HT requires the presence of a negative anode fall.[13] A more detailed analysis of boundary conditions for a quasi-1D model of a HT proposed the possibility of HT operation in the absence of the anode fall,[14] or with an anode fall that would be a function of thruster operating conditions, namely the discharge voltage and the mass flow rate. The internal probe measurements reported in Refs. 6-11 were mainly focused on characterizing the acceleration region in HT and, therefore, do not provide information on the anode fall behavior. However, they indicate thruster operation under conditions of a nearly zero[6-10] or a positive[11] anode fall. Also noteworthy is that the penetration of near-anode electrons into the anode cavity causes ionization of the working gas inside the anode [3].

In this letter, we report the results of non-disturbing measurements in the near-anode region of the 123 mm diameter laboratory Hall thruster[16] operating in the 0.2-2 kW power range. The plasma potential, plasma density and electron temperature are measured at a few millimeters from the anode with movable biased electrostatic probes of planar geometry and movable emissive probes. The anode sheath thickness, typically assumed to be several Debye lengths, $\lambda_D \sim 0.05\,\text{mm}$, is thus very small, which makes technically difficult the use of probe diagnostics inside the sheath. However, information about the sign and magnitude of the anode fall can be obtained through probing plasma in the presheath, at a few millimeters from the anode.

The laboratory thruster, the test facility and the slow-movable radial probes setup used in this study are described in Refs 16 and 17. The thruster was operated at xenon gas mass flow rates of $\dot{m} = 2-5\,\text{mg/s}$, and in the discharge voltage range of $V_d = 200-450\,\text{V}$. The magnetic field was kept constant ($B_{max} \approx 120$ Gauss along the channel median[16]). After the first set of experiments, with the biased probe, the dielectric coating, which appears on the anode surface in the course of thruster operation, was removed, and a second set of experiments, with the biased probe, was conducted. In the third set of experiments, after the anode was coated again, measurements of the plasma potential with a floating emissive probe and discharge current oscillations measurements were performed. After that, the anode was cleaned again and emissive probe measurements were repeated, along with current oscillation measurements. Experimental procedures (including data analysis) for biased and emissive probe measurements are described in Ref. 17. Oscillations of the discharge current in the 10-100 kHz wave band were measured using a low-inductance low-capacitance 1 Ω shunt, placed in the cathode circuit, and LeCroy LT264M digital oscilloscope connected through two 10:1 passive probes.

Fig. 1 shows $V_d$ vs. $I_d$ characteristics of the 123 mm Hall thruster with the clean and coated anodes (referred to as "Clean" and "Coat", respectively) for several mass flow rates. As can be seen, the characteristics are similar for two anodes. Fig. 2 shows the results of biased probe measurements in the near-anode region of the HT with the clean and coated anodes, for $\dot{m} = 5 \text{ mg/s}$. Zero potential is chosen at the anode. As can be seen from Fig. 2 (a), the plasma potential at 2 – 12 mm from the anode is higher than the anode potential in the case of the clean anode, and lower than the anode potential in the case of the coated anode. This indicates the presence of a negative, electron repelling, and a positive, electron attracting, anode fall, respectively.

The thruster anode serves also as a gas distributor. When the outer surface of the anode is coated with dielectric, the discharge current circuit supposedly closes at the inner side of the anode, where the plasma penetrates through the gas-injecting holes[3]. The combined cross-sectional area of the holes is significantly smaller than the conductive surface area of the clean anode. To achieve a discharge current similar to that of the clean anode, the coated anode necessitates additional ionization near and inside the anode. This could be the reason for the formation of a positive fall near the coated anode, with the magnitude of one to three times the electron temperature, $T_e$. Two experimental facts support this logic. Firstly, the formation of the anode coating in the thruster is associated with a visual effect: the gas-injecting holes start to glow brighter than the rest of the anode, with appearance of a jet-like structure from each hole. Secondly, in the N vs. $V_d$ characteristics presented in Fig. 1 (b), the density near the coated anode is almost twice as large as it is near the clean anode. Since ions in the positive anode fall are moving from the anode into the near-anode region, the high density might indicate enhanced ionization, which takes place near and inside the coated anode. The fact that the anode fall changes from negative to positive when the anode surface area is decreased is also well known for glow discharges.[18]

Fig. 2 (a) shows that the magnitude of the anode fall increases with the increase of the discharge voltage for the clean anode, and decreases for the coated anode. This correlates with the fact that the near-anode electron temperature increases with $V_d$ [Fig. 2 (b)]. At the same $I_d$, the increase of the near-anode $T_e$ can increase the negative anode fall and reduce the positive one, in three ways: a) by increasing ionization; b) by increasing the electron velocity toward the anode; and c) by increasing the electron bombardment of the anode, which, in the case of the coated anode, could lead to the partial removal of the dielectric coating and to the increase of the anode collecting surface area, AA. As can be seen from Fig. 2 (c), the positive fall near the coated anode decreases with the increase of the mass flow rate, at the same $V_d$. Ionization becomes more effective at larger $\dot{m}$, while the discharge current in Hall thrusters is limited by a magnetic field profile, which could result in the described behavior of the anode fall. However, the increase of AA due to the electron bombardment, which becomes more effective at larger $\dot{m}$, could also account for that fact.

The results of emissive probe measurements, presented in Fig. 3, corroborate the results of the biased probe measurements described above. It can be seen that the anode fall is, again, positive for the coated anode, and negative for the clean one. Also, the near-anode plasma potential generally increases with the increase of $V_d$ and $\dot{m}$. However, the tendencies are more complicated than those that follow from the biased probe measurements. This could be due to the poor spatial resolution typical for floating emissive probes.[19] Another reason could be that the thruster operating regimes were less stable in the experiments with the emissive probe.[16]

Interestingly, for the same condition of the anode surface, it was impossible to select thruster operating conditions [ $V_d$, $\dot{m}$, $B_r(z)$ ] in a *typical* operating range of the 12.3 cm HT in order to alter the sign of the anode fall: it was always positive for the coated anode and negative for the clean anode. Although the measured near-anode potential structure is essentially two-dimensional,[20] the above result holds for any radial location from near the inner to near the outer channel wall. However, as can be seen from Fig. 2 (c), at very large mass flow rates the sheath near the coated anode alters and becomes negative. Also, in a separate set of measurements in the acceleration region of the same 123 mm diameter thruster with a coated anode,[19] it was shown that at $\dot{m} = 3 \text{ mg/s}$ and $V_d \leq 300 \text{ V}$, the plasma potential at 21 mm from the anode lies below the anode potential, whereas at $V_d = 600 \text{ V}$ it lies several volts above. Taking into account that axial variations of the plasma potential in the near-anode region are relatively small [Fig. 2 (a)], the above must indicate altering of the anode sheath. The possible reason for this altering is, again, the increase of the near-anode electron temperature, which can reduce the positive anode fall in the three ways described above.

Discharge current oscillations were observed in the tens of kilohertz wave band, as shown in Fig. 4. The results of oscillation measurements indicate that thruster operation is more stable with the coated anode. The physical mechanism of this phenomenon is not yet understood. The typical oscillation frequencies increase with the discharge voltage from 7-10 kHz at $V_d \leq 200V$ to 15-25 kHz at $V_d \geq 300V$; for $\dot{m} = 5$ mg/s additional spectral maximum near 50 kHz appears at $V_d = 300V$. These oscillations may be critical for the thruster power processing system design and thruster integration with the satellite onboard circuitry.[21]

In summary, non-disturbing measurements in the near-anode region of the Hall thruster with biased and emissive probes showed for the first time the possibility of thruster operation with both a negative and a positive anode fall. It appears that the sign of the anode fall is essentially a function of the anode collecting surface area, rather than the thruster operating conditions. It was observed that the anode fall changes from positive to negative, at the same thruster operating conditions, when the dielectric coating, which appears on the anode surface in the course of operation, is removed.

The authors wish to thank D. Staack for his contribution to the preparation of the experiments, and A. Smirnov for fruitful discussions. This work was supported by the U.S. DOE under Contract No. DE-AC02-76CH03073.

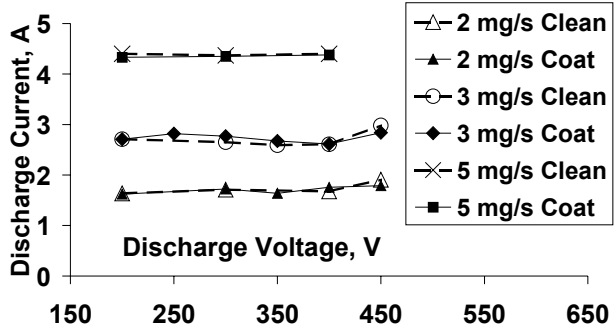

Fig. 1. $V_d$ vs. $I_d$ characteristics of the 12.3 cm HT with the clean and coated anodes for several mass flow rates.

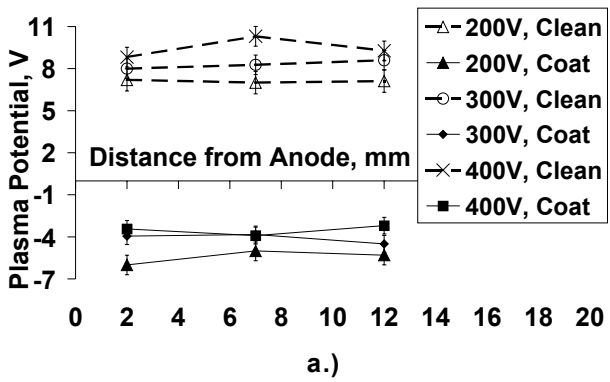

a.)

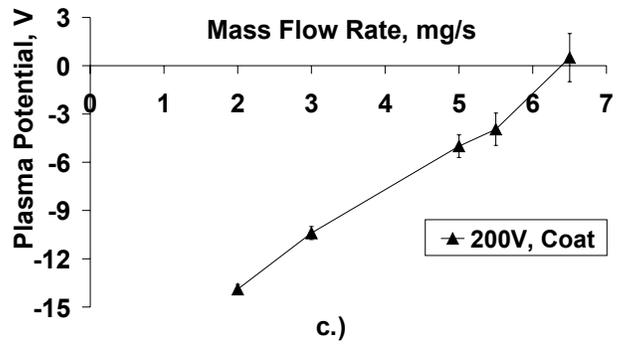

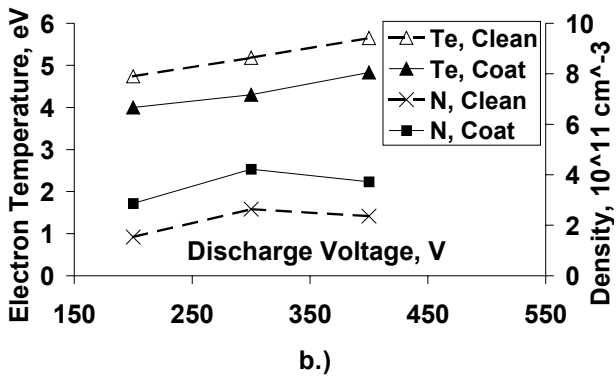

b.)

Fig. 2. Results of biased probe measurements in the near-anode region of the 12.3 cm HT. At the channel median: R = 49 mm from the thruster axis.

a.) With the clean and coated anodes. For $\dot{m} = 5\ mg/s$ and several discharge voltages.

b.) With the clean and coated anodes for $\dot{m} = 5\ mg/s$. At 7 mm from the anode.

c.) With the coated anode for $V_d = 200\ V$. At 7 mm from the anode.

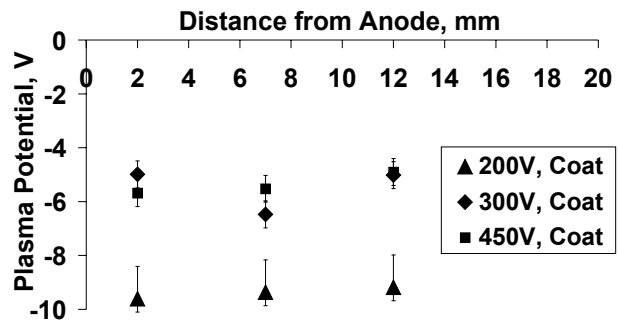

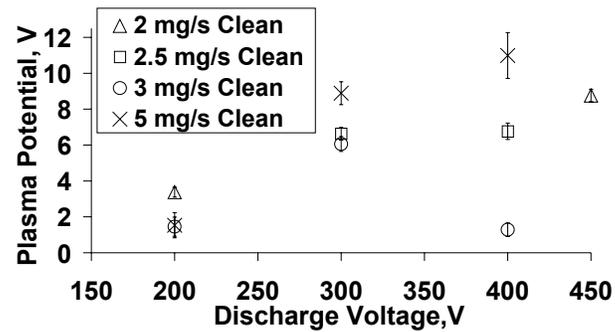

Fig. 3. Results of emissive probe measurements in the near-anode region of the 12.3 cm HT. At the channel median: R = 49 mm from the thruster axis.

a.) With the coated anode. For $\dot{m} = 3\ mg/s$ and several discharge voltages.

b.) With the clean anode. For several $\dot{m}$ and $V_d$. At 7 mm from the anode.

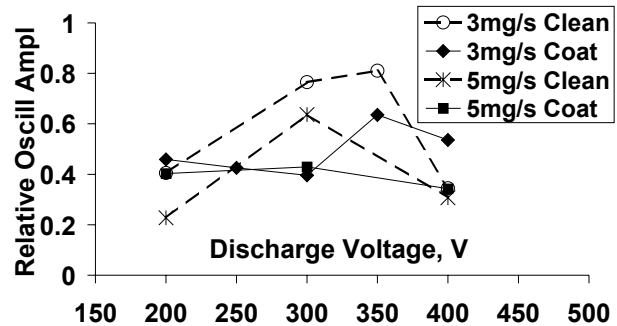

Fig. 4. Results of discharge current oscillations measurements in the 12.3 cm HT with the clean and coated anodes for the typical operating conditions.

REFERENCES

1. A. I. Morozov and V. V Savelyev, in *Reviews of Plasma physics*, edited by B. B. Kadomtsev and V. D. Shafranov (Kluwer, Dordrecht, 2000), Vol. 21.

2. G. W. Butler, J. L. Yuen, S. O. Tverdokhlebov, A. V. Semenkin and R. S. Jankovsky, Proceedings of the 36$^{th}$ Joint Propulsion Conference and Exhibit, July 2000, Huntsville, AL, AIAA Paper 2000-3254.

3. I. V. Melikov, Sov. Phys. Tech. Phys. **19**, 35 (1974)

4. Y. Raitses, J. Ashkenazy and M. Guelman, J. Prop. Power **14**, 247 (1998).

5. A. M. Bishaev and V. Kim, Sov. Phys. Tech. Phys. **23**, 1055 (1978).

6. G. Guerrini, C. Michaut, M. Dudeck, A. N. Vesselovzorov and M. Bacal, Proceedings of the 25$^{th}$ International Electric Propulsion Conference, Aug 1997, Cleveland, OH, IEPC Paper 1997-053.

7. J. M. Haas and A. D. Gallimore, Phys. Plasmas **8**, 652 (2001).

8. N. B. Meezan, W. A. Hargus, Jr. and M. A. Cappelli, Phys. Rev. E **63**, Art. No. 026410 (2001).

9. Y. Raitses, M. Keidar, D. Staack, N. J. Fisch, J. Appl. Phys. 92, 4906 (2002)

10. N. Z. Warner, J. J. Szabo and M. Martinez-Sanchez, Proceedings of the 28$^{th}$ International Electric Propulsion Conference, March 2003, Toulouse, France, IEPC Paper 2003-082.

11. A. I. Morozov, Yu. V. Esinchuk, G. N. Tilinin, A. V. Trofimov, Yu. A. Sharov and G. Ya. Shchepkin, Sov. Phys. Tech. Phys. **17**, 38 (1972).

12. Y. Raitses, L. A. Dorf, A. A. Litvak, and N. J. Fisch, J. Appl. Phys. 88, 1263 (2000)

13. E. Ahedo, P. Martinez-Cerezo, M. Martinez-Sanches, Phys. Plasmas **8**, 3058 (2001)

14. Dorf, V. Semenov, Y. Raitses and N. J. Fisch, Proceedings of the 38$^{th}$ Joint Propulsion Conference and Exhibit, July 2002, Indianapolis, IN, AIAA Paper 2002-4246.

15. M. Keidar, I. Boyd and I. Beilis, Proceedings of the 38th Joint Propulsion Conference and Exhibit, July 2002, Indianapolis, IN, AIAA Paper 2002-4107.L.

16. Y. Raitses, D. Staack, A. Dunaevsky, L. Dorf and N. J. Fisch, Proceedings of the 28$^{th}$ International Electric Propulsion Conference, March 2003, Toulouse, France, IEPC Paper 2003-0139.

17. L. Dorf, Y. Raitses, N. J. Fisch, Proceedings of the 28$^{th}$ International Electric Propulsion Conference, March 2003, Toulouse, France, IEPC Paper 2003-0157. Submitted for publication in Rev. Sci. Inst. (2003).

18. B. N. Kliarfeld and N. A. Neretina, Sov. Phys. Tech. Phys. **3**, 271 (1958).

19. Y. Raitses, D. Staack, L. Dorf and N. J. Fisch, Proceedings of the 39$^{th}$ Joint Propulsion Conference and Exhibit, July 2003, Huntsville, AL, AIAA Paper 2003-5153.

20. L. Dorf, Y. Raitses and N. J. Fisch, to be submitted.

21. N. Gascon, C. Perot, G. Bonhomme, X. Caron, S. Bechu, P. Lasgorceix, B. Izrar, and M. Dudeck, Proceedings of the 35$^{th}$ Joint Propulsion Conference and Exhibit, June 1999, Los Angeles, CA, AIAA Paper 1999-2427.